\documentclass[aip, jcp, showpacs, superscriptaddress, preprint ,11pt, longbibliography]{revtex4-1}
\usepackage{dcolumn,amsmath}
\usepackage{graphicx}
\usepackage{multirow}
\usepackage{wrapfig}
\usepackage{comment}
\usepackage{subcaption}
\newcommand{\wn}{cm$^{-1}$ }
\begin{document}

\title{Frequency measurements and self-broadening of sub-Doppler transitions in the $v_1+v_3$ band of C$_2$H$_2$ }

\author{Sylvestre\  Twagirayezu}\email{stwagirayezu@lamar.edu}\thanks{Present Address: Department of Chemistry \& Biochemistry, Lamar University, Beaumont, TX 77710}
\affiliation{Division of Chemistry, Department of Energy and Photon Sciences, Brookhaven National Laboratory, Upton, NY 11973-5000, USA}
\author{Gregory\ E.\ Hall}\email{gehall@bnl.gov}
\affiliation{Division of Chemistry, Department of Energy and Photon Sciences, Brookhaven National Laboratory, Upton, NY 11973-5000, USA}
\author{Trevor\ J.\ Sears}\email{trevor.sears@stonybrook.edu}
\affiliation{Division of Chemistry, Department of Energy and Photon Sciences, Brookhaven National Laboratory, Upton, NY 11973-5000, USA}
\affiliation{Chemistry Department, Stony Brook University, Stony Brook, NY 11794-3400, USA}

\begin{abstract}
Frequency comb-referenced measurements of sub-Doppler laser saturation dip absorption lines in the $v_1+v_3$ band of acetylene near $1.5\,\mu \mathrm{m}$ are reported.  These measurements include transitions involving higher rotational levels than previously frequency measured in this band.  The accuracy of the measured frequencies is typically better than 10 kHz.  Measurements of the observed sub-Doppler line widths as a function of pressure showed that the self pressure-broadening coefficients are about 3.5 times larger than those derived from conventional pressure broadening of unsaturated Doppler-limited spectra.  This is attributed to the contribution of velocity-changing collisions to the total dephasing rate in the low pressure sub-Doppler measurements.  At higher pressures, when the homogeneous broadening becomes comparable to the typical Doppler shift per elastic collision, the velocity changing collisions cease to contribute significantly to the incremental pressure broadening.  A  time-dependent soft collision model is developed to illustrate the transition between low and high pressure regimes of sub-Doppler pressure-broadening.
\end{abstract}

\date{\today}

\maketitle

\section{Introduction}
\label{Intro}
The $v_1+v_3$ band of acetylene, C$_2$H$_2$, is a strong vibrational combination band in the 1.5$\mu$m region of the spectrum.  It is a parallel band with a simple P- $(\Delta J = -1)$ and R- $(\Delta J = +1)$ branch rotational structure and has long been used as a secondary wavelength, and more recently\cite{Madej} frequency, standard.  Techniques for the routine frequency-measurement of optical spectra have opened the way to far more precise determinations of line positions and shapes\cite{Cich2013,Forthomme2015} than were previously available.  We have previously frequency-measured hot band lines in this spectrum for the purposes of calibrating Doppler- and pressure-broadened measurements.\cite{Twagirayezu2015}  These measurements were made using laser saturation dip spectroscopy with a fiber laser-amplified \textit{c.w.} extended cavity diode laser referenced to a frequency comb as the  source.

Here, we report rest frequency measurements for some lines in the spectrum that have not previously been frequency measured.\cite{Madej} During the frequency measurements, it was noted that the width of the saturation resonance varied with pressure, in a similar way to that seen in a conventional absorption spectrum.  The pressure and power dependence of the saturation dips have been measured and analyzed. The pressure broadening coefficients are typically about 3.5 times larger than those previously reported from conventional, Doppler-limited, spectroscopic measurements at much higher pressures.  Previous measurements of sub-Doppler saturation dip spectra in atoms such as Xe\cite{Cahuzac1978,Borde1977} and others discussed in section \ref{disc} also report self-pressure broadening coefficients larger than seen in Doppler broadened spectra.  Multiple observations, see for example references,\cite{Bagaev1972,Barger1969,Bagaev1979} of sub-Doppler spectra of methane in the 3.39 $\mu$m lasing transition have also been reported and these also exhibit larger self-pressure-broadening coefficients. Measurements of sub-Doppler CO$_2$ vibration-rotation transitions\cite{Meyer1975, Vasilenko1977,Kochanov1977} also show larger self-broadening than observed in Doppler-limited spectra

The interpretation of these and other similar observations has generally focussed on the effects of velocity-changing collisions. Physically, this is reasonable as sub-Doppler spectroscopy, in which a single velocity group of molecules is selected, might be expected to be more sensitive to weak, large cross section, collisions compared to Doppler-limited spectra where the line width is determined by the ensemble average of all velocities in the sample.  In the spirit of these ideas, we describe a simple collisional model based on a soft collision model\cite{Galatry1961, Rautian1967} and the interplay between damping and dephasing, which mimics the observations.

\section{Experimental}
\label{experiment}
Sub-Doppler spectra of selected ro-vibrational lines in the $v_1+v_3$ band of acetylene were measured by the transmission of an external cavity diode laser beam through a resonant cavity containing low pressures of acetylene.  The probe laser beam was locked to a cavity resonance using the Pound-Drever-Hall method,\cite{Black2001} with a DC-300 kHz error correction signal added directly to the diode injection current. The resonant cavity consisted of two highly-reflective mirrors, each mounted on a PZT piezo-electric actuator. One actuator with a slow response, but longer travel was used for coarse positioning of the cavity modes, the second with short travel, but higher frequency response executed an 830 Hz dither in addition to a stepwise ramp scan, generated by a second feedback loop that controlled the offset frequency between the probe laser and a reference tooth of the frequency comb.  With both feedback loops locked, the repetition rate of the frequency comb was stepped to change the average optical frequency of the synchronously dithered probe laser and sample cavity transmission maximum through the sub-Doppler resonance.   Data were recorded using an optical frequency step size of approximately 70 kHz through the center of each line, with steps three times larger elsewhere, to accelerate the data acquisition.  Each data point was typically averaged for 3s with a 300ms  output time constant for the lock-in amplifier.  A programmed delay was included after each frequency step to allow the system to stabilize before data acquisition began. More complete details have been previously reported.\cite{Twagirayezu2015,Twagirayezu2016}

As detailed below, the data analysis was carried out assuming a Voigt-type profile to account for a combination of transit-time and pressure-broadening effects.  The (narrow) transit-time contribution to the observations is most important for the lower pressure measurements and an analysis of the effect of changing the width of the Gaussian component is included in Supplementary Information for this paper.

We found the line shape modeling required a pressure-independent Lorentzian component to the line shape function, that could not be attributed to the effect of wavelength modulation broadening.\cite{Axner2001, Westberg2012}   Instead, this residual broadening is mostly due to fluctuations of the probe laser frequency derived from mechanical and acoustical noise sources.  The average laser frequency was accurately maintained by reference to the frequency comb, but the bandwidth of the phase lock loop holding the absolute comb frequency was slow enough that measured data points were an average over a Lorentzian distribution of frequencies centered about the exact center one. The 830 Hz component of the fluctuations was demodulated by the detection system.  We have modeled the effect quantitatively by assuming a Lorentzian distribution of frequency noise on top of 1-f wavelength modulation of a Voigt line shape. The convolution of two Lorentzian functions, \textit{i.e.} frequency noise plus a pressure-dependent broadening, leads to a new Lorentzian function with width parameter, $\gamma$, equal to the sum of the widths of the two constituents.  The modeling also showed that he modulation-induced broadening was small, and its effects were confined to the Gaussian, transit-time, contribution to the Voigt line shape, implying the actual transit time broadening was somewhat smaller than derived below.

\section{Results}

\subsection{Pressure-dependent line profiles}
\begin{figure} [t]  %{1.0\textwidth}
 	\subcaptionbox{$5.2\times 10^{-3}$ mbar}[.45\linewidth]
    	{\includegraphics[width=.45\linewidth]{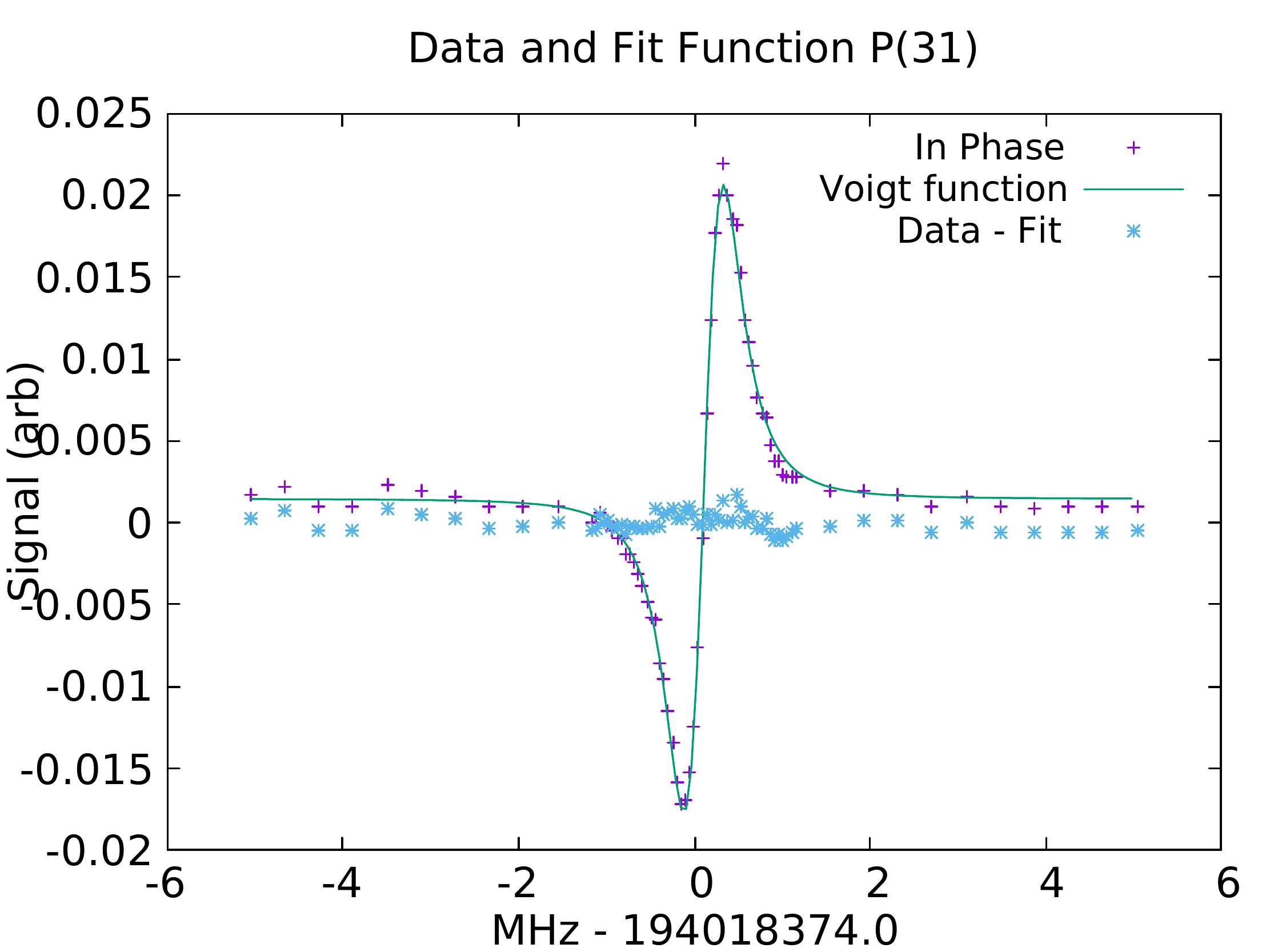}}
 	\subcaptionbox{$14.7\times 10^{-3}$ mbar}[.45\linewidth]{%
 		\includegraphics[width=.45\linewidth]{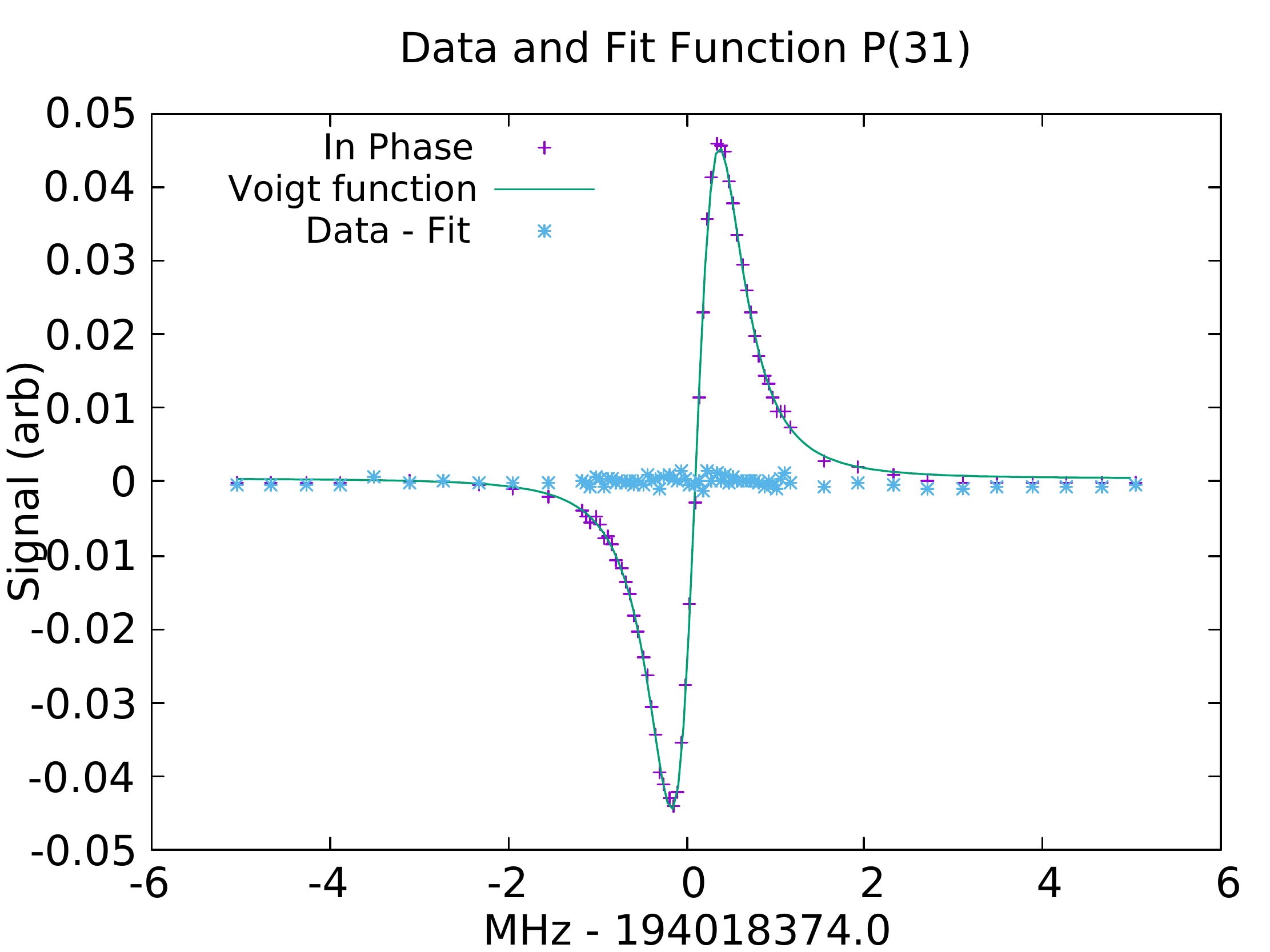}}
 	
 	\bigskip
 	
 	\subcaptionbox{$26.7\times 10^{-3}$ mbar}[.45\linewidth]{%
 		\includegraphics[width=.45\linewidth]{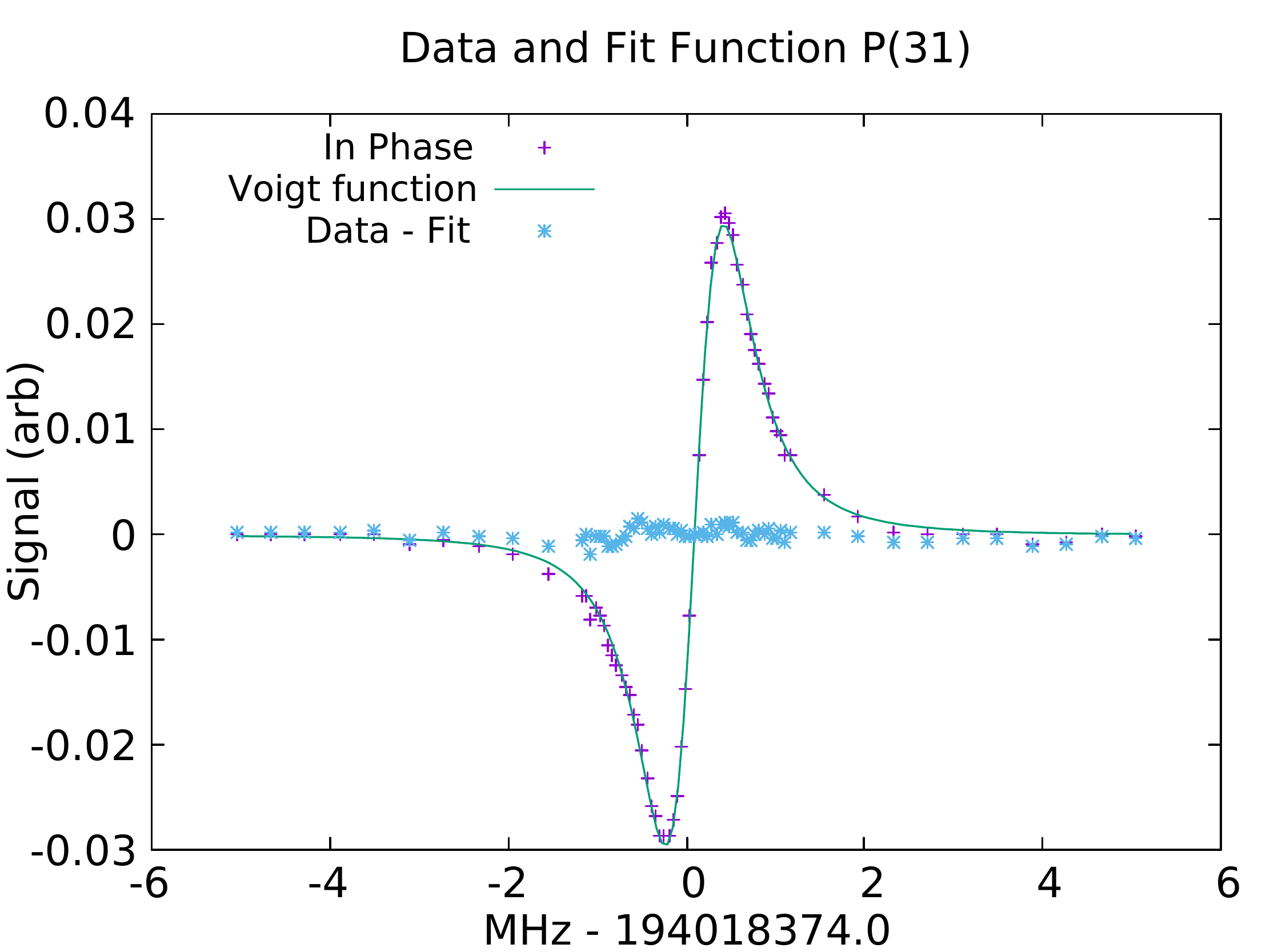}}
 	\subcaptionbox{$39.2\times 10^{-3}$ mbar}[.45\linewidth]{%
 		\includegraphics[width=.45\linewidth]{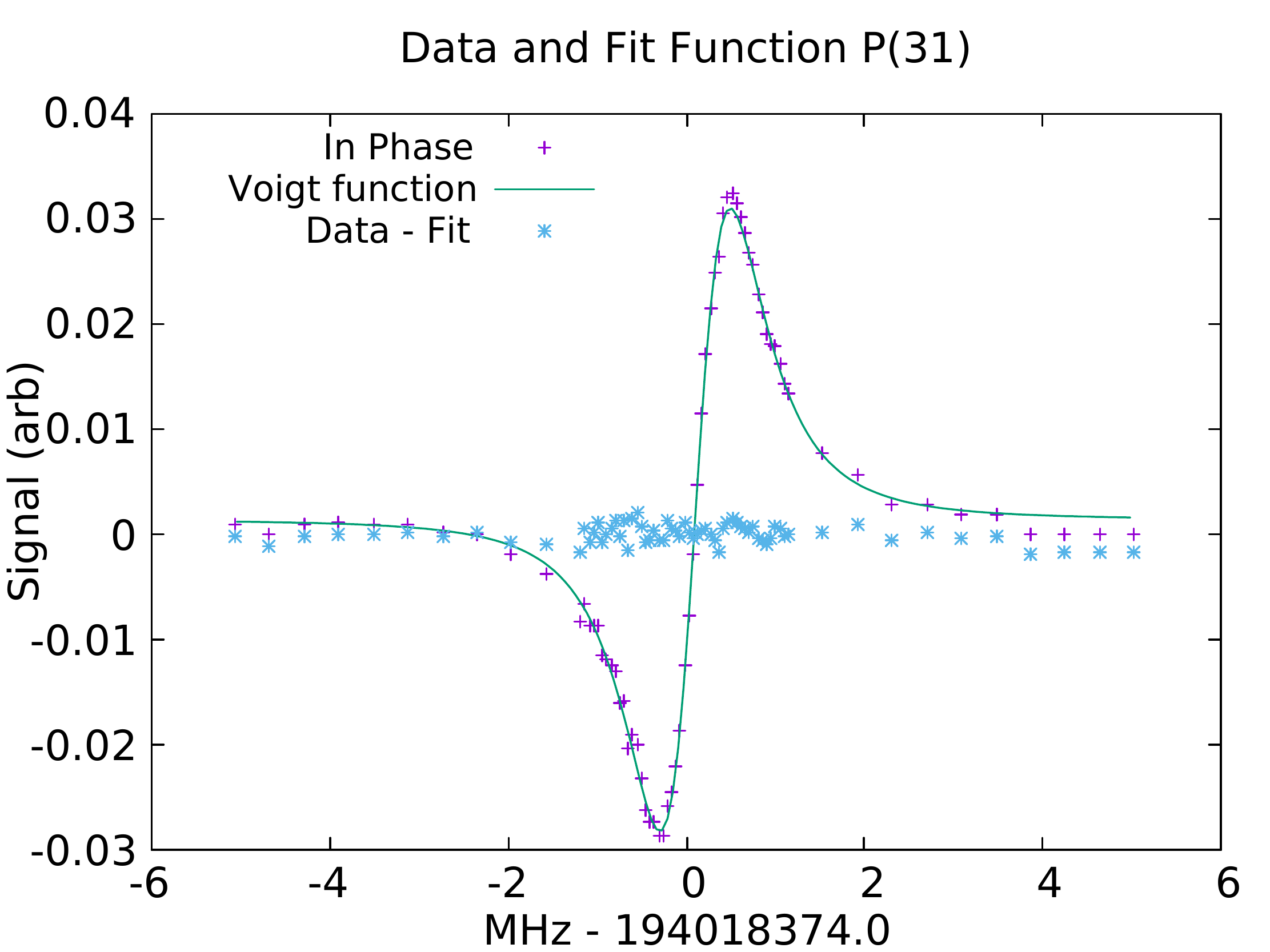}}
 	
 	\caption{\footnotesize{Saturation dip spectra for the P(31) line in the $v_1+v_3$ band of C$_2$H$_2$ at a series of sample pressures, shown in the subcaptions, 1 mbar=0.750062 Torr=100 Pa. Scans were recorded with a more dense sampling near line center. The increase in observed width with sample pressure is clear.  All  data were recorded at approximately 11 mW intracavity power. The line shapes were fit to a derivative of a Voigt profile consisting of a fixed Gaussian transit-time, and an adjustable Lorentzian component. }}
 	\label{PressBroadg}
 \end{figure}

 Sub-Doppler resonances for the P(31) line at several pressures are shown in Figure \ref{PressBroadg}.  In a low intensity limit, where the power broadening is negligible, the observed saturation dip line shape can be treated as the convolution of a Gaussian component due to transit-time broadening   \cite{Demtroeder, Borde1974, Hall1974, Borde1976, Ma1999} and a Lorentzian component due to multiple damping and dephasing mechanisms.  The transit-time broadening depends on the laser beam size and the distribution of transverse velocities in the molecular sample.  Most models of the effect just include the most probable Maxwell-Boltzmann velocity, $u_p=\sqrt(\frac{2k_BT}{m})$, although a more realistic description of the interaction would include the fact the slower molecules will contribute differently to faster ones\cite{Ma1999} and a description of the 2-D velocity distribution perpendicular to the laser axis.\cite{Dupre2018}  The convolution of the Gaussian and Lorentzian gives the well-known Voigt profile.  With an estimated beam waist radius in the cavity of $\approx 0.5$ mm and assuming the expression for transit-time broadening from Bord\'{e} and Hall,\cite{Borde1974, Hall1974, Borde1976} $\Delta \nu_{t_t} \approx \frac{u_p}{8w_0}$ we estimate a  Gaussian half-width-half-maximum (HWHM) of 0.112 MHz.  In fact, an initial  analysis of the P(31) data showed the Gaussian contribution to be constant over the pressure range studied, at approximately 0.105 MHz HWHM,  which is very close to the estimate.  We note that other derived approximate forms for $\Delta \nu_{t_t}$ such as Demtroeder's\cite{Demtroeder} $\frac{\sqrt{2ln2}}{2\pi}\frac{u_p}{ w_0}$ do not agree as closely with the experiment. Since the data for transitions other than P(31) were less extensive and generally of poorer signal-to-noise, in all subsequent analysis the Gaussian contribution to the measurements was assumed to be constant and fixed to 0.105 MHz HWHM.
 
\begin{figure} [h]
 	\includegraphics[width=0.45\textwidth] {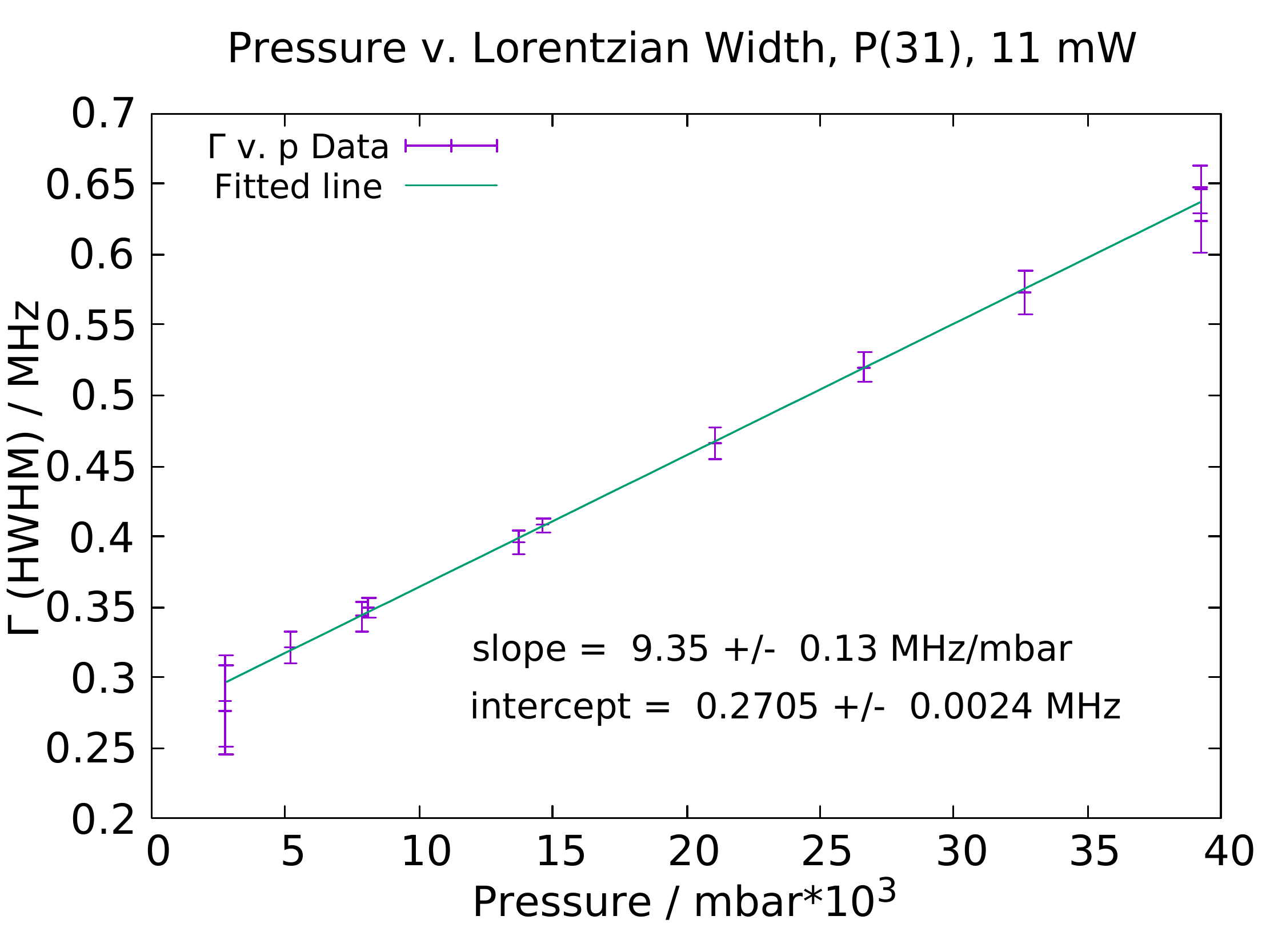}
 	\caption{\footnotesize{Plot of Lorentzian width $\Gamma$ for the P(31) saturation dip spectra in a Voigt fit with Gaussian HWHM fixed at 0.105 MHz.  The values are well described by a linear model.  The error bars on the points are three standard deviations of the fitted widths for the individual measurements. }}
 	\label{WidthvPress}
 \end{figure}

  Voigt fits to the pressure-dependent P(31) line profiles with fixed Gaussian HWHM of 0.105 MHz lead to the Lorentzian broadening as plotted in Figure \ref{WidthvPress}.  The intercept in Figure \ref{WidthvPress} is the extrapolated zero-pressure Lorentzian width, which corresponds to residual broadening due mainly to the mechanisms discussed above. The natural (fluorescence) lifetime for the upper level, J = 30 in the $v_1+v_3$ level can be estimated from known Einstein A-factors.\cite{Hitran2012}   It is approximately 50 Hz, too small to contribute significantly to the line widths. Power broadening was characterized in separate experiments in which data for P(31) were recorded at a pressure of $14.3 \times 10^{-3}$ mbar at a series of intracavity laser powers. Over the range 11-40 mW, the HWHM increased linearly with a slope of $\gamma_{pwr}=3.38(33)$ kHz/mW, suggesting the power broadening contribution to the spectra in Figure \ref{PressBroadg} is no more than 37 kHz.  Hence, the instrumental broadening due to the effect discussed in section \ref{experiment} is of the order of 230 kHz HWHM.
  
 The slope of the line in Figure \ref{WidthvPress}, \textit{i.e.} 9.35(13) MHz/mbar, or 12.46(17) MHz/Torr, is $\gamma_{self}$(HWHM): the self-broadening coefficient for the sub-Doppler P(31) transition at a one-way intracavity power of 11 mW.  The cavity-defined, counter-propagating, single laser beam geometry in the present measurements means that they may be compared directly to $\gamma_{self}(Doppler)$ derived from single-pass measurements of Doppler-broadened gaseous samples.  The sub-Doppler $\gamma_{self}$ determined here is about 3.5 times larger than  $\gamma_{self}(Doppler)$ for the same line.  Forward- and backward-going beams address separate velocity groups, each lifetime broadened. The convolution of a tunable probe with a fixed-frequency depletion would lead to feature that is double the width of either one alone.\cite{Barger1969} However, this effect is exactly compensated in a single beam cavity measurement, since the relative Doppler detuning of the forward and backward beams changes at twice the scan rate.  There was some confusion on this topic in the early saturation dip literature, \cite{Chebotayev1975} when comparing pressure broadening of sub-Doppler features and Doppler-broadened spectra at higher pressures.
 
 A series of saturation dip spectra at pressures from 2.9 to 29.4 mTorr ((3.9 - 39.2)$\times 10^{-3}$ mbar) for the P(31) - P(37) lines in the $v_1 + v_3$ band spectrum of acetylene were measured and analyzed. Table \ref{measureddata} summarizes the self pressure-broadening coefficients for sub-Doppler resonances in these transitions, along with literature values of the pressure broadening coefficients derived from Doppler-broadened spectra. All sub-Doppler coeffcients are significantly larger than their Doppler-broadened counterparts and we also note that the change in $\gamma_{self}$ measured here ($\approx$4\%) over this limited range of rotational quantum numbers is significantly smaller than for conventional Doppler-limited spectra, where there is a 15\% reduction from P(31) to P(35).  These observations imply that different mechanisms are active in the two types of experimental measurements and we discuss this below. There is much current interest in differences between self pressure-broadening in ortho- and para-acetylene.\cite{Iwakuni2016, Lehmann2017} Unfortunately, the estimated error in the self-broadening coefficient for the only para- line for which pressure broadening has been measured here, P(34), is too large to distinguish small ortho-/para- differences.

\subsection{Rest frequencies}

 The rest frequencies determined for the observed rotational transitions are also summarized in Table \ref{measureddata}.  The cited frequency uncertainties are the statistical (1$\sigma$) errors from multiple measurements of each line at various pressures between approximately 3 and 40$\times 10^{-3}$ mbar. We noted a slight tendency towards a negative pressure-frequency shift coefficient for the P(31) data, $\Delta_0 = -0.34(16)$ MHz/mbar with one standard deviation in parenthesis.  The estimated zero-pressure intercept from the fit of the P(31) data from this analysis was 194018374.106(3) MHz, \textit{i.e.} slightly larger than quoted in Table \ref{measureddata}, but within the combined estimated 1-$\sigma$ errors.  For P(11), we previously determined $\Delta_0=-0.4238(20)$ MHz/mbar from extensive Doppler-limited data.\cite{Forthomme2015}  If the present rotational transitions had a similar pressure-dependent shift coefficient, the low-pressure shifts would be close to the accuracy and precision limits of our current measurements, leading to no conclusive comparison of high and low pressure shifting coefficients.
 
 \begin{table}[ht]
 	\caption{Frequencies and broadening parameters of selected transitions in the $v_1 + v_3$ band of C$_2$H$_2$.}
 	\label{measureddata}
 	\scriptsize
 	\begin{tabular}{|c|c|c|c|}
 		\hline
 		\multirow{2}{*}{Transition}  & Frequency\footnotemark[1] & $\gamma_{self}$\footnotemark[2]  &  $\gamma_{self}(Doppler)$\footnotemark[3] \\
 		&  MHz  & MHz/mbar & MHz/mbar  \\
 		\hline
 		P(31) & 194018374.101(3)\footnotemark[4]   &  9.35(13) & 2.65   \\
 		P(32) & 193924561.231(6)  &               &  2.56 \\
 		P(33) & 193830012.907(3)  &  9.49(37) &  2.45  \\
 		P(34) & 193734752.121(10)  & 7.3(3.1) &  2.35 \\
 		P(35) & 193638764.411(6)   & 9.41(40) &  2.24  \\
 		P(36) & 193542055.096(30)  &  &  2.13  \\
 		P(37) & 193444626.334(20)  &  &  2.02 \\
 		\hline
 		\hline
 	\end{tabular}
 	\footnotesize{
 		\footnotetext[1] {Measured rest frequency from sub-Doppler spectroscopy (this work).}
 		\footnotetext[2] {Sub-Doppler self-broadening coefficient (this work).  }
 		\footnotetext[3] {Doppler self-broadening coefficient, average of literature values.\cite{Jacquemart2002, Kusaba2001}}
 		\footnotetext[4] {Uncertainties in parenthesis are one standard deviation in units of the last quoted significant figures.}
 	}
 \end{table}
 
 There have been relatively few direct comparisons of self-pressure shifts for the same transitions between Doppler-limited  and sub-Doppler resonances.  Recent very high precision measurements on sub-Doppler lines in the (3-0) band of carbon monoxide\cite{Wang2017} found an upper bound to the pressure shift of a few tenths kHz/mbar. A significantly larger CO self-shift of 0.16 to 0.31 MHz/Torr (120-230 kHz/mbar) was reported in Doppler and pressure broadened FTIR spectra.\cite{Swann2002}  Sub-Doppler measurements on a methane transition at 3.39 $\mu m$\cite{Barger1969} found an unmeasurably small (75 $\pm$ 150 Hz/Torr) self-pressure shift coefficient at low pressures, while conventional IR absorption methane spectra have been reported to have much larger self-pressure shifts of -0.017 cm$^{-1}$ atm$^{-1}$ (-670  kHz/Torr) for transitions in the 6000 \wn region.\cite{Lyulin2011}  Other sub-Doppler IR measurements of carbon monoxide\cite{Wappelhorst1997} and water\cite{Chen2018} similarly led to upper bounds smaller than the high-pressure self-pressure shift. Precise measurements of saturation dip spectra of ethylene in the 10 $\mu$m region\cite{Tochitsky1998} and iodine\cite{Goncharov1990} in the visible showed detectable, small self-shifts, but with a nonlinear pressure dependence.  Self-collisional frequency shift coefficients for sub-Doppler HD overtone lines\cite{Cozijn2018} were, however, found to be much larger at -900 kHz/mbar, some 10$\times$ larger than found in Doppler-broadened spectra of the normal isotopomer, H$_2$.\cite{Tan2014} Overall then, there is little or no pressure-induced shift seen in most of the sub-Doppler experiments and theoretical consideration of this phenomenon in the literature is sparse or non-existent.

\subsection{Time-Dependent Collisional Model}
The influence of velocity-changing collisions (VCCs) on the widths of low-pressure sub-Doppler line profiles has been investigated by numerical evaluation of model electric field correlation functions, which detail the damping and dephasing that are responsible for the broadening of a resonant absorption, in the spirit of Galatry\cite{Galatry1961} and Rautian and Sobel'man.\cite{Rautian1967}

The line profile is given by the real Fourier transform of the correlation function $\Phi(\tau)$:
\begin{equation}\label{equ1}
 I(\omega-\omega_0)\,=\,\frac{1}{\pi}\Re \int_0^{\infty}e^{i(\omega-\omega_0)\tau}\Phi(\tau) d\tau.
\end{equation}
The total correlation function can be approximated as the product of an exponential damping function $\Phi_{\gamma}(\tau)$, a fly-out or transit time damping function $\Phi_{t_{t}}(\tau)$, and a velocity-dependent dephasing function $\Phi_{v}(\tau)$
\begin{align}
 \Phi(\tau) \;\; &= \, \Phi_{\gamma}(\tau) \label{Phivtau}
 \Phi_{t_{t}}(\tau) \Phi_{v}(\tau)\\
 \Phi_{\gamma}(\tau)\, &= \, e^{-\gamma \tau}\\
 \Phi_{t_{t}}(\tau)\, &= \,e^{-(\tau /t_{t})^2} \label{Phitt} \\
 \Phi_{v}(\tau)\, &= \, \left< e^{-i {\bf k \cdot r} (\tau)}\right> \label{average}
\end{align}
with
\begin{equation}
 {\bf r}(\tau)\,=\, \int_t^{t+\tau} {\bf v} (t^{\prime})dt^{\prime}.
\end{equation} \\

The damping rate $\gamma$ is primarily a pressure-dependent inelastic damping rate but may also include a zero-pressure radiative contribution.  The transit time contribution is approximated as a Gaussian function of time, with typical collisionless transit time, $t_{t}$.   The Doppler effects due to a static or fluctuating velocity distribution are contained in the correlation function $\Phi_{v}(\tau)$, where the angle brackets in Eq.(\ref{average}) denote an ensemble average.  The motion of the absorber along the propagation direction of the light introduces a time-dependent phase, ${\bf k\cdot r}(\tau)$.   For  time $\tau$ shorter than the time interval between collisions, the velocity ${\bf v}$ is constant, and ${\bf r}(\tau)={\bf v} \tau$.  The phase, ${\bf k\cdot r}(\tau)$, is then a linear function of time, corresponding to a fixed Doppler frequency offset of $v/\lambda$, with $v$ the component of velocity along the wavevector ${\bf k}$.   For a thermal sample with a low collision rate, the ensemble average of velocities gives a Gaussian correlation function as the low pressure limit of $\Phi_{v}(\tau)$. The Fourier transform gives the usual Gaussian Doppler line profile, $I(\omega - \omega_0)$, for which the transit time damping is typically negligible compared to the rapid dephasing from multiple Doppler shifts.  Including an additional exponential damping, through $\Phi_{\gamma}(\tau)$, leads to a Voigt profile.  As the collision rate increases with pressure, each absorber may experience more than one Doppler-shifted frequency before suffering an inelastic collision or leaving the interaction region.  In this case, the correlation function $\Phi_v(\tau)$ will dephase more slowly than the inhomogeneous average over multiple Doppler shifts, characterizing the narrowing associated with the velocity diffusion.  The confinement at higher pressures makes Eq. (\ref{Phitt}) an increasingly inaccurate representation of transit time contributions, although a pressure-dependent increase in effective transit time only renders this slow damping an even more negligible contribution to the total correlation function at higher pressures.

For application to sub-Doppler line profiles, the inhomogeneous averaging over the thermal velocity distribution is left out, and only the resonantly selected velocity class of molecules with a specific initial Doppler shift is considered, while retaining the same treatment of velocity diffusion,  inelastic loss of population, and  the transit time.

The central point of the model is to depict the variable effect of velocity changing collisions with pressure, as the pressure-dependent homogeneous line width changes from small to large compared to the typical Doppler shifts accompanying an elastic collision.  Relevant previous discussions of these questions can be found, for example, in several  references from the early days of saturation spectroscopy.\cite{Bagaev1972,Meyer1975,Vasilenko1977,Cahuzac1978,Goncharov1990}

Neglected in this model are any pressure shifts, collision effects in the excited state, properly treated with a density matrix approach, power broadening and the variation of the saturation parameter on the pressure, and any effects of radially variable intensities, wavefront curvature and recoil effects, all important in more quantitative treatments.\cite{Borde1977, Ciurylo_2001, Wehr2006, Dupre2018}  The finite duration of the collisions is neglected by assuming a negligible phase shift during the collision.  The speed-dependence of collision rates has also been neglected, as has any explicit treatment of residual broadening due to instrumental effects discussed above.

It is worth noting that the velocity change associated with inelastic collisions, while contributing substantially to the conventional diffusion constant and the overall translational thermalization rate, is irrelevant to pressure broadening, since the state change immediately removes the molecule from the near-resonantly driven ensemble under consideration.  It is only the differential cross sections for elastic (state-preserving) collisions that need be considered in the evaluation of $\Phi_v(\tau)$. When expressed as a laboratory-frame collision kernel, the forward peak in the elastic differential cross section leads to a cusp-like distribution of velocity change along a viewing direction.\cite{Ho1986, Gibble1991, McGuyer2012}  In the present numerical illustration, the collision kernel is approximated as a symmetric exponential function of the velocity change along the viewing direction:
\begin{equation}\label{equ7}
 A(v,v^{\prime})\, \propto \, e^{-|v-v^{\prime}|/\Delta_{vcc}}
\end{equation}
with $\Delta_{vcc}$  the average absolute value of the velocity change following an elastic collision. This form of a collision kernel does not have the correct asymptotic form required to converge at long times to a Maxwell-Boltzmann velocity distribution, but it can quite accurately represent the first few elastic collisions with small total displacements in velocity space compared to the thermal velocity.   Inelastic collisions truncate the sequence of elastic velocity changing collision long before thermalization, so that this failure to satisfy asymptotic reversibility is not a serious problem.   Realistic calculations of elastic differential cross sections peak at scattering angles less than 2 degrees, which, when combined with the kinematic averaging of center of mass to laboratory frame velocity changes find typical Doppler shifts of a few MHz per collision for C$_2$H$_2$ transitions in the near infrared at room temperature.

\begin{figure} [t]
	\includegraphics[width=0.5\textwidth] {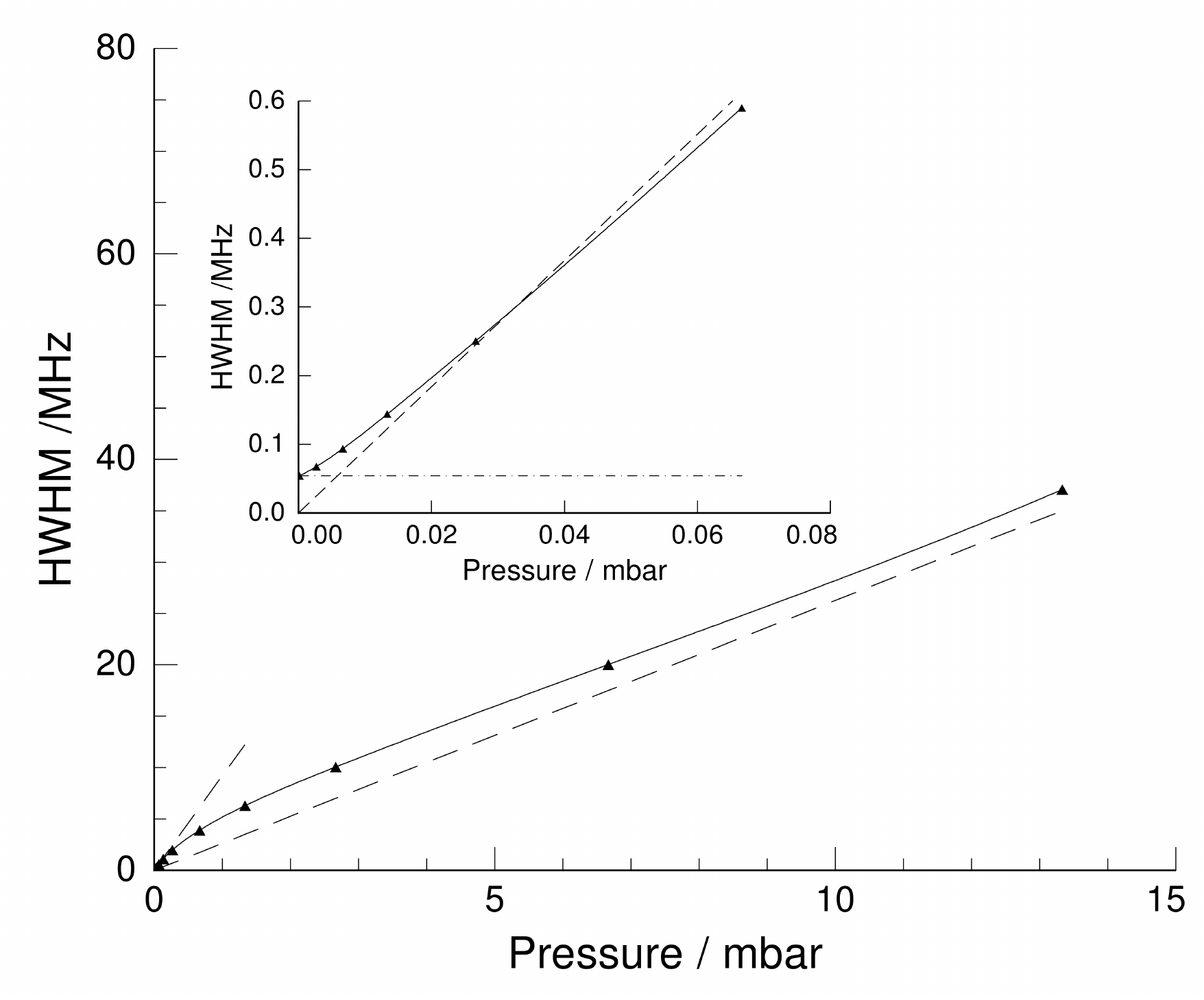}
	\caption{\footnotesize{  Simulated pressure dependence of sub-Doppler line broadening.  At pressures above 3 mbar, the HWHM increases with a slope characteristic of the inelastic rate. The steeper dashed line at lower pressures has a slope characteristic of the total elastic + inelastic rate.  The low pressure inset shows the transit time limit as a horizontal dashed line, and the steeper sloped line from the main figure.}}
	\label{HWHMvPress}
\end{figure}

\begin{figure} [h]
    \includegraphics[width=0.3\textwidth]
    {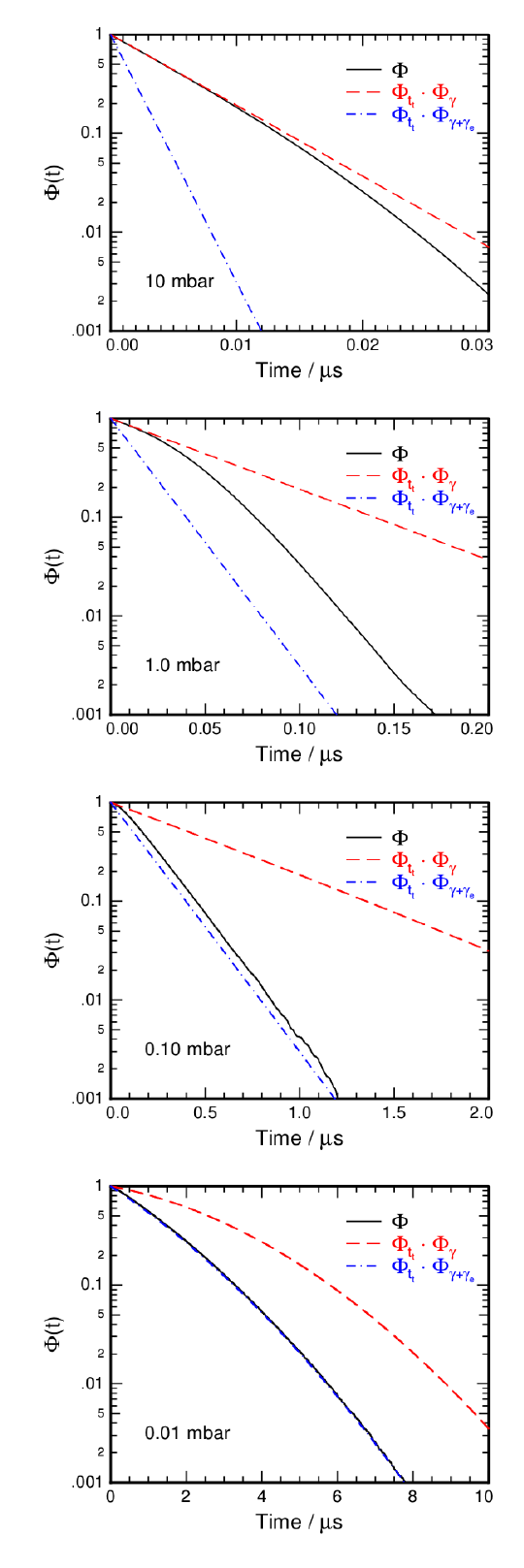}  \caption{\footnotesize{Correlation functions computed from 0.01 to 10 mbar. Solid black lines are the full simulation, a product of   $\Phi_{\gamma}(\tau),\,\Phi_{t_{t}}(\tau),\,\mathrm{and}\, \Phi_{v}(\tau)$  At high pressures, the computed correlation function approaches the red dashed line, which neglects all velocity changing elastic collisions.  At low pressures, the computed correlation function approaches the blue dot-dashed line, representing the limit where all elastic collisions contribute to the effective damping. }}
\label{PhivTime}
\end{figure}

With these simplifications, a numerical simulation based on Eq. (\ref{equ1}-\ref{equ7}) for the pressure-dependent sub-Doppler line profiles can be made with only three relevant parameters: the ratio of elastic to inelastic collision rates $\gamma_e / \gamma$,  the typical Doppler shift per elastic collision $\Delta_{vcc}/ \lambda$, and a transit time $t_{t}$.  In numerical calculations designed to mimic the conditions of the present acetylene measurements, the correlation function is evaluated with 100 ps time steps from 0 to 50 $\mu$s.  The transit time contribution $\Phi_{t_{t}}(\tau)$ is evaluated with $t_{t} = 5 \mu$s.  The inelastic contribution  $\Phi_{\gamma}(\tau)$ is evaluated with the damping rate $\gamma\,=\,2\pi p\cdot2.65\times 10^6$ $\mu$s$^{-1}$ mbar$^{-1}$, with pressure $p$ in mbar, chosen to match the experimental 2.65 MHz/mbar high-pressure self-broadening coefficient (HWHM)  for a J = 31 line of C$_2$H$_2$.\cite{Jacquemart2002, Kusaba2001}
The velocity-dependent correlation function $\Phi_v(\tau)$  is evaluated by a Monte Carlo procedure with one parameter for the elastic to inelastic collision ratio, $\gamma_e / \gamma$, and another to parameterize the typical magnitude of the velocity change per collision, $\Delta_{vcc}$.   For each time array in the ensemble average, the time array is partitioned into a sequence of collisionless intervals, $\{\Delta t_{i}\}$, with a random sample of times between elastic collisions generated from an exponential distribution with a mean value $< \Delta t> \,=\,1/ \gamma_e$. For each molecular trajectory in the ensemble average, then, the phase factor ${\bf k \cdot r}(\tau)$  in Eq (\ref{average}) is a continuous, piecewise linear function of time, with a local slope in each collisionless time interval equal to the current detuning $v_i / \lambda$. The molecule follows a random walk in velocity $\{v_i\}$  starting at zero displacement from the laser-selected initial value.  The velocity steps $v_{i+1} - v_i$   are sampled from the collision kernel, $A(v,v^{\prime})$ of Eq (\ref{equ7}), with $\Delta_{vcc} / \lambda$  set to 3 MHz.  For the illustrations in Figures (\ref{HWHMvPress}) and (\ref{PhivTime}), $\gamma_e / \gamma$ is set to 2.5.  Demonstration code implementing this algorithm is available in the Supplemental Material for this paper in the form of a stand-alone Windows executable file and a LabView 2011 library.

Figure \ref{HWHMvPress} shows the resulting trend in HWHM as a function of pressure, based on the Fourier transform of the simulated correlation function, using the model parameters above.   Two limiting pressure regimes can be seen, with local slopes corresponding to the total collision rate at low pressure, and just the inelastic collision rate at high pressures.  The transition between high and low pressure regimes occurs when the total pressure broadening is comparable to the Doppler shift per collision.

Figure \ref{PhivTime} shows correlation functions at several pressures.  In the low pressure region, the curvature in the plots of $log \, \Phi(\tau)$ comes from the pressure-independent quadratic transit time contribution, which becomes negligible at higher pressures, when compared to the increasingly fast collision-induced damping.  Throughout the pressure range below 0.05 mbar, where the present acetylene measurements were made, the ensemble average of the elastic collisions results in nearly full dephasing at the total collision rate, $\gamma_e + \gamma$.  This is consistent with good fits to a Voigt line shape, and a linear pressure dependence of the Lorentzian broadening, as observed experimentally.  The slope of HWHM vs. pressure in this region is the total collision rate/2$\pi$.   The  line shapes begin to deviate from Voigt forms by 0.1 mbar as the  increasingly fast inelastic damping rate begins to compete with the dephasing between different velocity groups. Eventually, at high enough pressures, the velocity changing collisions have no further incremental effect on the pressure broadening. Our measurement of a sub-Doppler pressure broadening with a slope 3.5 times the value obtained by conventional spectroscopy in Doppler-broadened gaseous samples is consistent with an elastic collision rate 2.5 times larger than the inelastic contribution.   The linearity of the Lorentzian fitting parameter up to 0.05 mbar is consistent with a typical Doppler shift per elastic collision of at least 2 MHz, according to this simple model.

\section{Discussion}\label{disc}

Using saturation dip laser absorption spectroscopy, we have frequency measured the positions of a number of rotational lines in the $v_1 + v_3$ combination band of acetylene.  For the stronger transitions, we were able to determine the effect of self-collisional broadening of the line profiles as the sample pressure was increased. The broadening coefficients determined in the present experiment, see Table \ref{measureddata}, are all $3.5-4 \times$ those previously reported in conventional, Doppler-broadened spectroscopic measurements of the same transitions.  To account for and understand this observation, we have developed a sub-Doppler soft collision model which takes into account the effect of large cross-section, VCCs in the sample.  The correlation function $\Phi_v(\tau)$, Equ. (\ref{average}), actually accounts for several types of velocity-dependent effects in line shapes. In the absence of VCCs at low pressure it gives the normal Gaussian Doppler broadening when averaged over the thermal velocity distribution.  Including the effects of VCCs modifies this ensemble average such that extreme velocity groups in the ensemble tend to become more similar with collisions and slow down dephasing. This is a way of describing Dicke narrowing.\cite{Galatry1961, Farrow1987}  In the sub-Doppler case, the initial ensemble of interest is velocity-selected and VCCs broaden the velocity distribution and cause a corresponding line broadening.  The particular model detailed here with a symmetric small-angle scattering kernel would not show Dicke narrowing however, as this requires significant excursions in velocity space for each molecule prior to inelastic scattering and a kernel with asymmetric probabilities for decreasing and increasing velocity changes in collisions.
  
Other reported measurements of molecular sub-Doppler lines have also found 2-4$\times$ greater pressure broadening coefficients compared to measurements of Doppler-broadened lines.\cite{Barger1969,Bagaev1972,Wappelhorst1997,Ma1999,Hald2011,Chen2018,Meyer1975,Vasilenko1977,Kochanov1977}  The most detailed studies of the effects of VCCs on saturated absorption profiles have been in atoms such as Xe,\cite{Cahuzac1978,Cahuzac1979} and Rb with  He, Ne, Ar and Xe collision partners.\cite{Gibble1991} The extremely strong atomic transitions permit measurements over a wide range of pressures.  These studies show the same qualitative behavior as depicted in our model, although pressure-broadening in them is primarily driven by phase interrupting atomic collisions, rather than the rotationally inelastic collisions that dominate molecular systems. At low pressures, the narrow sub-Doppler saturated resonance is observed to have a larger Lorentzian broadening component than calculated, based on the pressure broadening coefficient measured in conventional absorption studies.  Line widths are also found to increase with a non-linear pressure dependence: linear at low pressures followed by a region of downward curvature, approaching a high pressure linear region with a slope near the value measured for the Doppler-broadened gas-phase absorption data.
 
 Similar results were seen in methane\cite{Bagaev1972,Bagaev1979} in early measurements of the 3.39 $\mu\mathrm{m}$ lasing transition.   The direct effect of VCCs with increasing pressure can also be seen in the formation of a broad pedestal under the narrow saturation feature under appropriate conditions.\cite{Bagaev1979}  Measurements of vibration-rotation transitions in CO$_2$ found a $\gamma_{self}$ for the saturation features that was 2.7$\times$ larger than the Doppler-broadened quantity.\cite{Vasilenko1977,Kochanov1977}  A similar result for CO$_2$ was reported earlier by Meyer et al.\cite{Meyer1975} and incorrectly attributed to a combination of transit time broadening and saturation effects.
 
Recent measurements of the self-broadening of sub-Doppler spectra of CH$_3$F in the 3 $\mu$m region\cite{Okabayashi2015} are of interest to compare with the present results.  Okabayashi et al.\cite{Okabayashi2015} report $\gamma_{self}$=9.6(1.3) MHz/Torr for the $^rQ(7,4)$ transition, converted to the HWHM quantity, which compares to a value of 10.7(0.5) previously reported by Cartlidge and Butcher\cite{Cartlidge1990} from linear spectroscopy in the Doppler-limited regime.  Published data for ammonia\cite{Triki2012,Mattick1973, Mattick1976,Hitran2012} also find no significant differences between the self-broadening coefficients for Doppler or sub-Doppler broadening.  This implies that, for these molecules, the effects of VCCs in the saturation dip experiment are minimal, contrary to the case of acetylene, methane, CO$_2$ or ethylene, above.  The obvious difference is the presence of the dipole moment in CH$_3$F and NH$_3$, causing even long-range collisions to be dominated by dipole-dipole interactions more effective at causing state- or phase-changing transitions than in molecules or atoms without a dipole moment.
 
These observations are in accord with measurements and modeling of the decay of coherent transients reported by Berman et al.\cite{Berman1975} Their work investigated the effects of velocity changing collisions by measurement of photon echo signals following rapid Stark switching of the molecular transition frequency out of resonance with the probe laser frequency. The conclusions were that, in CH$_3$F, the decay of a Doppler hole created by a narrow band laser is dominated by hard, inelastic, collisions.  Further, the measurements were interpreted to show that phase-changing collisions in the absence of state-changing ones, were also of low probability, so that the T$_1$ and T$_2$ lifetimes were the same. Much more recent work by Rubtsova et al.\cite{Ledovskikh2010,Rubtsova2011} confirmed this result.
Stark-shifting measurements are only possible on molecules possessing a dipole moment and degenerate or nearly degenerate levels of opposite parity, so cannot be applied to acetylene, for example, where the present results suggest VCCs play a much more significant role.
  
In related work by the Schwendeman group,\cite{Shin1991b,Song1992,Soriano1998} collisions in CH$_3$F were probed by elegant sub-Doppler IR-IR double resonance methods.  They observed collisional transfer of velocity spikes characteristic of resonant rotational energy transfer.  After such a collisional rotational energy exchange, a new population of a nearby rotational state is formed with the narrow velocity distribution of the initially pumped one, and a corresponding component of the  initially tagged level appears to be fully thermalized translationally in a single collision.  This special type of collision between indistinguishable particles may of course also be considered, rather than the complete translational thermalization of the tagged quantum state, as a small-angle soft collision, but with an exchange of rotational state labels. In either view, these special energy-elastic self-collisions will contribute to pressure broadening, despite having no observable effect on the energy transfer rates. It is the relative importance of these collisions that is at issue in the current interest in possible ortho/para differences in the pressure broadening of acetylene.\cite{Iwakuni2016, Lehmann2017}
 
It would be interesting to measure coherent transients in acetylene using a rapid frequency-switching, or transient double resonance, experiment to compare with the present interpretation of the sub-Doppler saturation dip data.  Precise understanding of the various contributions to sub-Doppler line profiles are crucial to the interpretation of spectroscopic experiments that attempt to test fundamental theory\cite{Cozijn2018, Tao2018} while information on the rate of VCCs is required for the modeling precise experimental measurements of pressure-broadened lines recorded under Doppler-broadened conditions relevant to remote sensing\cite{Ngo2013} and Doppler thermometry.\cite{fischer2018}

\section{Supplementary Material}
Supplementary material consisting of a) compressed files containing LabView$^{\textcircled{c}}$ executable code and source code libraries to permit readers to implement the model for computing the effects of velocity-changing collisions on sub-Doppler lineshapes described in this paper and b) a pdf document showing the effect of a larger width Gaussian transit-time broadening on the fits to a Voigt profile, may be obtained from the publisher.

\section{Acknowledgements}
We gratefully acknowledge helpful discussions with Dr Patrick Dupr\'{e} (Universit\'{e} Littoral C\^{o}te d'Opal) and Dr Benoit Darqui\'{e}, (Universit\'{e} Paris-13).   Work at Brookhaven National Laboratory was carried out under Contract No. DE-SC0012704 with the U.S. Department of Energy, Office of Science, and supported by its Division of Chemical Sciences, Geosciences and Biosciences within the Office of Basic Energy Sciences.
\footnotesize

%\bibliography{Refs-C2H2-0119}

\end{document}